\newcommand{\bigq}{\qquad\qquad\quad}

\documentclass{elsart}

\begin{document}

\begin{frontmatter}

\title{PopRatio: A program to calculate atomic level populations in astrophysical plasmas}
\author[IAG]{A.I. Silva\thanksref{e-mail1}} and
\author[IAG]{S.M. Viegas\thanksref{e-mail2}}
\address[IAG]{Instituto Astron\^omico e Geof\'\i sico, Universidade de S\~ao Paulo,
	Av. Miguel St\'efano, 4200, 04301-904 S\~ao Paulo SP, Brazil}
\thanks[e-mail1]{E-mail: ignacioalex@yahoo.com .}
\thanks[e-mail2]{E-mail: viegas@iagusp.usp.br .}

\begin{abstract}

We describe a Fortran 90 program to calculate population ratios of atomic levels.
The program solves the equations of statistical equilibrium considering all possible
bound-bound processes: spontaneous, collisional or radiation induced (the later either
directly or by fluorescence). There is no limit on the number of levels or in the number
of processes that may be taken into account. The program may find a wide range of
applicability in astronomical problems, such as interpreting fine-structure absorption lines
or collisionally excited emission lines (such as coronal emission lines) in the spectra
of several objects, and also in calculating the cooling rates due to collisional excitation.

PACS: 95.30.Dr; 98.38.Bn; 98.58.Bz

\end{abstract}

\begin{keyword}

Level populations, statistical equilibrium equations, fine structure lines,
collisionally excited lines, coronal lines, cooling rates.

\end{keyword}

\end{frontmatter}

{\bf PROGRAM SUMMARY}

{\em Title of program:} PopRatio

{\em Catalogue identifier:}

{\em Program obtainable from:} http://www.iagusp.usp.br/\~{}alexsilv/popratio

{\em Computer for which the program is designed and others on which it has been tested:}

{\em Computer:} Intel Pentium Pro 200 MHz PC

{\em Installation:} Instituto Astron\^omico e Geof\'\i sico, Universidade de S\~ao Paulo.

{\em Operating system under which the program has been tested:} DOS, Windows 95.

{\em Programming language used:} Fortran 90.

{\em Memory required to execute with typical data:} \\
2,055 kbytes

{\em No. of bits in a word:} 32

{\em No. of processors used:} One

{\em Has the code been vectorised or parallelized?:} No


{\em No. of bytes in distributed program, including test data, etc.:} \\
981 058

{\em Distribution format:}

{\em Keywords:} \\
Level populations, statistical equilibrium equations, fine structure lines,
collisionally excited lines, coronal lines, cooling rates.


{\em Nature of physical problem:}  \\
The purpose of this program it to calculate population ratios of atomic levels.
The excited level population ratios may be used to infer the physical conditions from
observed fine-structure absorption lines or collisionally excited emission lines
(such as coronal emission lines). Another possible use
is in calculating cooling rates due to collisional excitation of low-lying levels.

{\em Method of Solution:} \\
The necessary atomic data is read from a separate input file. Next, the rates for all
the bound-bound processes involved are evaluated and the system of equilibrium equations is
solved. Several processes may be taken into account: spontaneous, collisions with an arbitrary
number of particles, excitation or stimulated emission induced by radiation fields
(either directly or by fluorescence). Built in radiation fields provided are:
a black body (such as the cosmic microwave background radiation), the UV radiation
field of the Galaxy, the UV background of all QSOs and the hot halo model radiation field.
Moreover, an arbitrary user-defined radiation field may also be included.

{\em Restrictions on the complexity of the problem:} \\
None. The code can handle an arbitrary number of processes and levels.
The required memory is dynamically allocated in run time.

{\em Typical running time:} \\
The following table gives the running times (in seconds) of the testcases provided
in section \ref{section:testcases}:

\begin{tabular}{ l l l l}
\hline
&                & System library & Distributed libraries \\
\hline
& testcase \#1   & 0.06 & 0.06 \\
& testcase \#2   & 0.06 & 0.10 \\
& testcase \#3   & 0.06 & 0.06 \\
& testcase \#4   & 0.05 & 0.07 \\
& testcase \#5   & 0.12 & 0.14 \\
\hline
\end{tabular}

The listed values correspond to averages over 10 executions on a PC Pentium Pro 200 MHz.
The column labelled "System library" gives the running times in case the routines from the
MSIMSL library are used.
The figures given in the next column correspond to the distributed version of PopRatio,
that includes routines from the BLAS, LAPACK and PPPACK libraries. The testcases were run using
unoptimized BLAS routines.


{\bf LONG WRITE-UP}

\section{Introduction}

In most astrophysical environments, common atoms or ions are found primarily in their ground
states, with only a negligible fraction in excited levels.
However, if some excitation mechanism is present, then a small or even significant fraction of
atoms/ions will be found in their excited states. If the excitation mechanism is not so
intense to keep the atoms/ions in LTE conditions, then the amount of atoms/ions found in
excited states may be used to infer the physical conditions in their environments.

The fraction of excited atoms/ions may be infered from the observed spectrum properties.
For example, the ratios of excited level populations may be deduced from the column
density ratios of fine structure absorption lines seen in absorption clouds towards QSOs
or in diffuse clouds towards brigth stars in the Galaxy \cite{Bahcall}.
They may also be infered from intensity ratios of collisionally excited emission lines
(such as coronal emission lines).

However, in order to trace the diagnosis curve giving the fraction of excited atoms/ions
as a function of the physical conditions in the medium, one must usually account for several
excitation mechanisms \cite{Smeding}. Moreover, some atoms/ions may have a complex enough
electronic structure to require the inclusion of several levels in the calculation,
such as iron.

This paper describes a code to calculate the population ratios of excited levels of a
given atom or ion accounting for an arbitrary number of levels and excitation mechanisms.

Our code may find a wide range of applicability in astronomical problems, such as infering
the physical conditions from fine-structure absorption line and collisionally
excited emission line
diagnosis. Another possible use is in the calculation of cooling rates due to collisional
excitation. We provide several examples in the testcases described bellow. 

\section{Method of Solution}

\label{section:theory}

\subsection{The equations of statistical equilibrium}

In order to calculate the level populations of a given
atom or ion, we make two basic assumptions :

\begin{enumerate}

\item The rates of processes involving ionization stages other than the atom or ion
being considered (such as direct photoionization or recombination, charge exchange reactions,
collisional ionization, etc.) are slow compared to bound-bound rates.

\item All transitions considered are optically thin.

\end{enumerate}

To calculate the population of some level $i$, we must take into account all possible
processes that will (de)populate it:

\begin{equation}
\frac{dn_i}{dt} = \sum \left( \mathrm{processes}\ \mathrm{that}\ \mathrm{populate} \right) - 
	\sum \left( \mathrm{processes}\ \mathrm{that}\ \mathrm{depopulate} \right), 
\end{equation}

where $n_i$ is the volume density of atoms or ions in level $i$.

Therefore, in steady state regime the sum over all processes that populate level $i$ will
be balanced by the sum over all processes that depopulate level $i$. Assuming that the two
conditions listed above are met, this can be written
(see, for instance, Rybicki and Lightman \cite{Rybicki}):

\begin{equation}
\sum_j n_j \left( A_{ji} + B_{ji}u_{ji} + \sum_k n^k q^k_{ji} \right) =
n_i \sum_j \left( A_{ij} + B_{ij}u_{ij} + \sum_k n^k q^k_{ij} \right) ,
\label{eq:sum}
\end{equation}

where we have considered all possible bound-bound processes, i.e., spontaneous, 
radiation-induced and collisionally-induced.
The lefthand side of eq. (\ref{eq:sum}) is the sum over all processes that populate level $i$
from the other levels $j$, whereas the righthand side is the sum over all processes that
depopulate level $i$ to levels $j$.

$A_{ij}$ is the transition probability of spontaneous decay from level $i$ to level $j$. For 
$i \leq j$, $A_{ij}=0$.

$B_{ij}$ are Einstein coefficients, related to the transition probabilities by:

\begin{eqnarray}
B_{ij} & = & \frac{1}{8 \pi h} \frac{A_{ij}}{{\left( E_i - E_j \right)}^3} \nonumber \\
B_{ji} & = & \frac{g_i}{g_j} B_{ij} ,
\label{eq:EinsteinB}
\end{eqnarray}

for $i>j$, and $B_{ii}=0$; $h$ is Plank's constant, $E_i$ is the energy of level
$i$ (expressed in ${\mathrm{cm}}^{-1}$) and $g_i$ is the statistical weight of level $i$.

$u_{ij}$ is the spectral energy density of the radiation field integrated along the line profile
$\phi_{\nu}$ of the transition from level $i$ to level $j$:

\begin{equation}
u_{ij} = u_{ji} = \int u_{\nu} \phi_{\nu} d \nu \simeq \int u_{\nu} \delta \left(
\nu - \nu_{ij} \right) d\nu = u_{\nu} \left( \nu_{ij} \right) ,
\end{equation}

with $u_{ii}=0$; $\nu_{ij}$ is the frequency of the transition and we have assumed that the radiation field
does not vary significantly along the line profile.

In eq. (\ref{eq:sum}) we have also considered the effect of collisions; $n^k$ is the volume
density of the particle inducing the transition, the main collision partners usually being
$k = e^- , p^+, H^0, {He}^0, H_2, ...$ , depending whether the medium is primarily ionized or
neutral.

$q^k_{ij}$ is the collision rate for the transition from level $i$ to level $j$ induced by
the collision partner $k$ .
These coefficients are the cross-sections for the related process $\sigma_{ij}$ 
convolved with a Maxwellian distribution of velocities $f \left( v \right)$, making
these quantities suitable for astrophysical applications
(see, for instance, Osterbrock \cite{Osterbrock}):

\begin{equation}
q_{ij} = \langle v \sigma \rangle = \int v \sigma \left( v \right) f \left( v \right) dv =
\frac {1}{{\left( kT \right)}^2} \sqrt {\frac{8kT}{\pi\mu}} \int_0^{\infty} \sigma_{ij} 
\left( \epsilon \right) \epsilon e^{-\frac{\epsilon}{kT}} d\epsilon ,
\label{eq:q}
\end{equation}

for the deexcitation rates ($i>j$); $k$ is Boltzmann's constant, $T$ is the kinetic temperature,
$\mu$ is the reduced mass of the system and $\epsilon$ is the collision partner's kinetic energy.

Excitation and deexcitation rates are related by the principle of detailed balance:

\begin{equation}
q_{ij} = \frac {g_j}{g_i} e^{-\frac{E_j - E_i}{kT}} q_{ji} ,
\label{eq:detailedbalance}
\end{equation}

with $q_{ii}=0$.
When the interaction is coulombian, as in collisions with electrons, it is convenient
to express the cross-section in terms of the {\em collision stregth\/} $\Omega_{ij}$, 
defined by:

\begin{equation}
\sigma_{ij} \left( \epsilon \right) = \frac {h^2}{8 \pi m \epsilon} 
\frac {\Omega \left( \epsilon \right)}{g_i} ,
\end{equation}

where m is the electron's mass. Substituting this in eq. (\ref{eq:q}) yields:

\begin{equation}
q^{e^-}_{ij} = \frac {h^2}{m g_i} \sqrt{\frac{1}{8{\pi}^3 \mu kT}}
\int^{\infty}_0 \Omega_{ij} \left( \epsilon \right) e^{-\frac{\epsilon}{kT}} 
d \left( \frac {\epsilon}{kT} \right) = \frac {8.629 \ {10}^{-6}}{\sqrt{T}} 
\frac {\gamma_{ij}}{g_i} ,
\label{eq:gamma}
\end{equation}

with $T$ expressed in K and $\gamma_{ij}$ is defined by the integral in eq. (\ref{eq:gamma}) 
and is called {\em Maxwellian-averaged collision stregth}. Typically $\gamma_{ij}$ is a slowly
varying function of T, of order unity. However, for neutral atoms it may vary for several
orders of magnitude.

These are the basic parameters needed to solve eq. (\ref{eq:sum}). If we consider our model ion
to be composed of $n$ levels, then we must solve a linear system of $n-1$ equations in order
to calculate the relative population ratios.

\subsection{Fluorescence}

\label{section:fluorescence}

Sometimes we have the situation in which the atom or ion is in one of its first $m$ 
lower-lying levels and then it is photoexcited to some higher-lying level $\lambda$ (e.g., by an
UV radiation field). Next it decays - either spontaneously or by stimulated emission - back to
some different level among the first $m$ levels. We call this process {\em fluorescence}.

In this case it is easy to eliminate the level $\lambda$ from the linear system of equations
(\ref{eq:sum}), reducing its order by one.

Initially, for notation purposes let us define:

\begin{equation}
K_{ij} \equiv B_{ij} u_{ij}
\end{equation}

The statistical equilibrium equation (\ref{eq:sum}) for some level $l$ belonging to the first
$m$ lower-lying levels is:

\begin{equation}
n_{\lambda} \left( A_{\lambda l} + K_{\lambda l} \right) + \cdots = n_l K_{l\lambda} + \cdots ,
\label{eq:eq1}
\end{equation}

where we have written only the terms involving the level $\lambda$. Whereas the equation for
level $\lambda$ will be:

\begin{equation}
\sum_{i=1}^m n_i K_{i\lambda} =
n_{\lambda} \sum_{i=1}^m \left( A_{\lambda i} + K_{\lambda i} \right) .
\label{eq:eq2}
\end{equation}

Solving for $n_{\lambda}$ in eq. (\ref{eq:eq2}), substituting in eq. (\ref{eq:eq1}) and
introducing the {\em indirect excitation rate\/} from level $i$ to level $j$ through level
$\lambda$ as:

\begin{equation}
\Gamma_{ij}^{\lambda} \equiv K_{i\lambda} \frac {A_{\lambda j}+K_{\lambda j}}
{\sum_{g=1}^m \left( A_{\lambda g} + K_{\lambda g} \right)} ,
\end{equation}

with $\Gamma^{\lambda}_{ii}=0$; we can now rewrite eq. (\ref{eq:eq1}) as,

\begin{equation}
\sum_{j=1}^m n_j \Gamma_{jl}^{\lambda} + \cdots = n_l \sum_{j=1}^m \Gamma_{lj}^{\lambda} +
 \cdots .
\label{eq:eq3}
\end{equation}

And we have eliminated the equation involving $n_{\lambda}$.
Extending this reasoning to eliminate a whole set of $\mu$ higher-lying levels is
straightforward.
One simply replaces the indirect excitation rates in eq. (\ref{eq:eq3}) by the corresponding
{\em total indirect excitation rates\/}:

\begin{equation}
\Gamma_{ij} \equiv \sum_{\mu} \Gamma_{ij}^{\mu} .
\end{equation}

\subsection{Solving the system of statistical equilibrium equations}

Let us now proceed to set the system of equations to calculate the relative population ratios.

For notation purposes, let us define:

\begin{equation}
Q_{ij} \equiv A_{ij} + K_{ij} + \Gamma_{ij} + \sum_k n^k q_{ij}^k ,
\end{equation}

so that eq. (\ref{eq:sum}) reads (taking into account the possible effect of fluorescence):

\begin{equation}
\sum_j n_j Q_{ji} = n_i \sum_j Q_{ij} ,
\end{equation}

or equivalently in matrix notation:

\begin{eqnarray}
P \cdot R = O \nonumber \\
P_{ii}    & = & -\sum_j Q_{ij} \nonumber \\
P_{ij}    & = & Q_{ji} \quad (i\neq j) \\
R_i       & = & n_i \nonumber \\
O_{ij}    & = & 0 \nonumber . 
\end{eqnarray}

After rewriting the system to have the solution expressed in terms of the populations
relative to the ground level $n_1$ we are left with:

\begin{eqnarray}
M \cdot X = I \nonumber \\
X_i & = & \frac{n_{i+1}}{n_1} \nonumber \\
I_i & = & -P_{i+1,1} = -Q_{1,i+1} \label{eq:lsystem} \\
M_{ij} & = & P_{i+1,j+1} = \cases{Q_{j+1,i+1}, & for $i\neq j$;
\cr -\sum_j Q_{i+1,j}, & for $i=j$. \cr} \nonumber
\end{eqnarray}

Eq. (\ref{eq:lsystem}) is the linear system of equations that is solved by PopRatio.

\section{The package PopRatio}

\subsection{The structure of PopRatio}

PopRatio is divided in four modules: PRECISION, PHYSMATH, FIELDINTENSITIES and POPRATIO.
Next, we describe each one in greater detail.

\subsubsection{PRECISION}
\label{section:Precision}

This module defines single/double precision kind parameters and sets the accuracy
to which the floating point operations are to be done throughout the entire package.
The user may change the precision of floating point operations by simply editing the WP
parameter:

{\bf integer}, {\bf parameter} :: WP = DP \quad ! working precision definition

The default value is set to correspond to the obsolescent Fortran 77's {\bf double precision}
data type.

\subsubsection{PHYSMATH}

This module contains definitions of $\pi$ and some physical constants,
as well as some mathematical routines.

The mathematical routines are:

\begin{itemize}
\item SWAP: subroutine that swaps any two integer numbers;
\hfill\break
\item POL: function that evaluates a polynomial at a given point,
\hfill\break
\item LINTERPOL: function that interpolates linearly a set of points.
\end{itemize}

\subsubsection{FIELDINTENSITIES}

This module contains routines that give the spectral energy densities (in erg/cm$^3$/Hz) 
of some particular radiation fields taken from the literature. The routines take as input
the wavelength in \AA\ at which the field is to be evaluated (as well as some additional
parameters required by each particular model) and interpolates linearly the table of
points using routine LINTERPOL.

The radiation fields currently available are:

\begin{itemize}
\item FGAL: this function returns the radiation field of the Galaxy, as given by Gondhalekar
et al. \cite{Gondhalekar};
\hfill\break
\item FUVB: this function returns the UV background radiation field of all QSOs, as given by
Madau et al. \cite{MHR}. It also takes as input the redshift at which the radiation field is
to be evaluated Z and a flag MODEL. MODEL=1 corresponds to a cosmological model with
$q_0=0$, MODEL=2 to one with $q_0=0.5$ and MODEL=3 corresponds to $q_0=0.5$ as revised by
\cite{MHR},
\hfill\break
\item FHOTHALO: this function returns the radiation field of the hot halo model proposed
by Viegas and Fria\c ca \cite{VF} to explain the source of ionization of Lyman Limit QSO
absorbtion line systems. It takes as input two flags: R and T. R is the distance
from the center of the galaxy at which the radiation field is to be evaluated, R=1 corresponds
to 10 kpc, R=2 to 30 kpc and R=3 to 100 kpc. T is the age of the galaxy, T=1 corresponds to
0.206 Gyr and T=2 to 0.3644 Gyr. 
\end{itemize}

\subsubsection{POPRATIO}

This is the main module that actually computes the rate coefficients for the processes
described in section \ref{section:theory} (after reading the related atomic data from a
separate input file) and then solves the system of equations of statistical equilibrium.

It begins defining a few useful constants to the evaluation of the rate coefficients and
three new types:

\begin{itemize}
\item TABLE: this type is a table of points, Y vs. X;
\hfill\break
\item TCOLLISIONALDATA: this type encloses all necessary information pertaining a certain
collisional process, such as the collision partner name, the collision rates, etc.,
\hfill\break
\item THLEVEL: this type encloses all necessary information pertaining some of the $\mu$
higher-lying levels described in section \ref{section:fluorescence}, such as the energy,
statistical weight, transition probabilities, radiative rates, etc.
\end{itemize}

This module also declares several global variables that are visible to the main program.
Table \ref{table:globalvariables} shows the correspondence between most important variables
and the mathematical quantities given in section \ref{section:theory}.

\begin{table}
\caption{Correspondence between global variables in module POPRATIO and mathematical quantities.}
\begin{tabular}{ l l l}
\hline
& Variable & Mathematical quantity\\
\hline
& E(i)   & $E_i$\\
& g(i)   & $g_i$\\
& A(i,j) & $A_{ij}$\\
& B(i,j) & $B_{ij}$\\
& u(i,j) & $u_{ij}$\\
& CD(k)\%q(i,j) & $q^k_{ij}$\\
& gam(i,j) & $\Gamma_{ij}$\\
& matrix(i,j) & $M_{ij}$\\
& indep(i) & $I_i$\\
& X(i) & $X_i$\\
\hline
\end{tabular}
\label{table:globalvariables}
\end{table}

We also have the following routines:

\begin{itemize}
\item SKIP\_COMMENTS: subroutine used by INITPOPRATIO to skip the comments from
the atomic data input file (see section \ref{section:inputfile} bellow);
\hfill\break
\item INITPOPRATIO: subroutine that reads in the atomic data values stored in the input file.
It takes as input a unit number for the input file (defined via an {\bf open} statement in the
main program), the number of levels NBLEVELS to be taken into account in the calculation, and
wavelength bounds to the radiation field intensities, LAMBDAI and LAMBDAF.
The last two arguments are optional parameters, and if supplied by the
user INITPOPRATIO will not load in any upper level connected to the ground
term by multiplets whose wavelengths lie outside of this range (the default
values are 0.0\_WP and {\bf huge}(1.0\_WP)).
INITPOPRATIO does not load
into RAM memory all the data stored in the input file, only the necessary space to take into
account the desired number of levels is allocated in run time. (See also section 
\ref{section:inputfile} below);
\hfill\break
\item FINISHPOPRATIO: subroutine that deallocates the dynamical variables allocated by
INITPOPRATIO;
\hfill\break
\item POPULATIONRATIO: this is the key subroutine that sets the rates for all the processes and
calculates the population ratios. It takes as input the kinetic temperature of the gas T (in K),
the redshift Z, a vector N containing the volume densities (in cm$^{-3}$) of the collision
partners entered in the atomic data input file and two optional parameters: BETA and F.

The redshift is used to set the temperature of the Cosmic Microwave Background Radiation (CMBR),
as given by the following relation predicted by the standard Big Bang cosmology
(see, for instance, Kolb and Turner \cite{KolbTurner}):

\begin{equation}
T=T_0 \left( 1+z \right),
\label{eq:TlawStandard}
\end{equation}

where $T_0=2.728\pm 0.002\ K$ is the current value of the CMBR temperature \cite{Fixsen,Smoot}.
The effect of the CMBR blackbody radiation field is then included in the energy densities:

\begin{equation}
u_{ij}= \rho \left( \nu_{ij},T \right) = \frac{8\pi h}{c^3} 
\frac{{\nu}^3_{ij}}{e^{\frac{h\nu_{ij}}{kT}}-1},
\label{eq:Planck}
\end{equation}

where c is the speed of light.
Alternatively, there are models that generalize the standard temperature law of the CMBR
(\ref{eq:TlawStandard}) by introducing a free parameter $\beta$ \cite{LimaAlexSueli}:

\begin{equation}
T=T_0 {\left( 1+z \right)}^{1-\beta},
\label{eq:Tlaw}
\end{equation}

where $\beta$ is within the range $0 \leq \beta \leq 1$.
If this parameter is not entered via the optional parameter BETA, it is assumed zero.
POPULATIONRATIO calls function TCMBR to evaluate the temperature of the CMBR, and next function
PLANCK to evaluate its contribution to the energy densities. If the user does not wish to
take the CMBR into account, the redshift Z should be entered with the value -1.0\_WP.

POPULATIONRATIO also adds to the energy densities the contribution from a radiation field
defined by the user in function URAD. The optional parameter F multiplies the radiation field
defined in URAD by a constant factor $f$. If F is not entered, it is assumed 1.

\hfill\break
\item LTEDEV: this function returns the module of the largest relative deviation
of the population of the lowest NL levels from the population expected from LTE
conditions characterized by a temperature T\_LTE (in K);

\hfill\break
\item URAD: this function returns the spectral energy density of a user-defined radiation field.
The user may create his/her own subroutine and then call it in the body of this function.

\hfill\break
\item PLANCK: this function returns the spectral energy density of a blackbody radiation field
(\ref{eq:Planck}). It first checks whether the argument of the exponential does not exceed
the natural logarithm of the largest number in the mashine representation in order to avoid
overflow (the spectral energy density is set to zero in case this happens).

\hfill\break
\item TCMBR: function that returns the temperature of the CMBR (\ref{eq:Tlaw});

\hfill\break
\item RATE\_EXTRAPOL: this function is called by POPULATIONRATIO whenever the kinetic
temperature T lies outside the range allowed by some table of points of collisional rates
entered in the atomic data input file (see section \ref{section:inputfile} below).
The user must supply a subroutine to extrapolate the table of points and modify this function
to call it, otherwise a warning message will be issued telling which particular table of points
needs to be extrapolated.
An analytical fitting to some transition may be implemented by entering an unphysical
temperature range in the input file, so that whatever value of T is entered, POPULATIONRATIO
will always call this function that may be used to call another function containing the fitting
(such as CI\_GAM and OI\_PROTON, see example in section \ref{section:inputfile} below);

\hfill\break
\item CI\_GAM: function that returns the Maxwellian averaged collision strengths 
for transitions - induced by collisions with electrons - involving the lowest five levels
of C$^0$ (based upon analytical fittings taken from the literature 
\cite{Johnson,PequignotAldrovandi}),

\hfill\break
\item OI\_PROTON: function that returns the deexcitation rates by collisions with protons for
transitions involving the lowest three levels of O$^0$ (based upon analytical fittings taken
from the literature \cite{Pequignot}). 

\end{itemize}

\subsubsection{Other packages used by PopRatio}

PopRatio uses several routines from the BLAS \cite{BLAS}, LAPACK \cite{LAPACK} and
PPPACK \cite{PPPACK} libraries.
The routines needed are distributed together with PopRatio in case the user does not
have them installed in his/her system.

We use the Fortran 90 interface to the LAPACK routines described by Dongarra et al.
\cite{LAPACK90}. To call the PPPACK routines we have written a new module - SPLINE -
containing Fortran 90 interfaces to the routines and a new function that calls them: SPLINE3.
The PPPACK routines were slightly modificated in order to support both real and
double precision arguments.

\subsection{The atomic data input file}
\label{section:inputfile}

The user must supply all necessary atomic data used by PopRatio in a separate input file,
that will be read by INITPOPRATIO.
We provide input files containing atomic data pertaining five atoms/ions of interest in
fine structure absorption line diagnosis of QSO absorbers: C$^0$, C$^+$, Si$^+$, O$^0$
and Fe$^+$. Although in these files we list all the references in the literature where the
atomic data were taken, we do not discuss the particular models adopted here,
since this will be the subject of a forthcoming paper (Silva and Viegas, in preparation).

Next we describe the general syntax of the input file. As an example we take the file OI.dat,
that contains the atomic data of neutral oxygen.

The atomic data information is distributed along several blocks, each one starting with the "\#"
character. The user is free to insert any comments before this symbol. The first block:

\# \break
OI

is the name of the atom/ion being considered. This string is stored in the global variable 
SPECIES\_NAME. It may be up to 10 characters long. The next blocks of data are:

\# \break
3

this is the number of levels that belong to the ground term. In atomic oxygen these levels are
$^3$P$_{2,1,0}$. PopRatio will take into account fluorescence for transitions among these levels.
This value is stored in the global variable NGROUND.

\# \break
5

this is the maximum number of levels that may be taken into account. The user must make sure
to enter enough data to allow PopRatio to consider this number of levels. If INITPOPRATIO is
called with a value for NBLEVELS higher than this, a warning message will be issued.

\begin{tabbing}
\bigq \=\qquad \=\qquad \=\qquad  \kill
\# \\
0.0         \> 5  \> 1  \> 2s2 2p4 3P2 \\
158.265     \> 3  \> 2  \> 2s2 2p4 3P1 \\
226.977     \> 1  \> 3  \> 2s2 2p4 3P0 \\
15867.862   \> 5  \> 4  \> 2s2 2p4 1D2 \\
33792.583   \> 1  \> 5  \> 2s2 2p4 1S0
\end{tabbing}

these are the values for the energies and statistical weights for the levels. The first column
is the energy relative to the ground level (in cm$^{-1}$), and the second is the statistical
weight for the level. The last two columns numbering and naming the levels are not read by
INITPOPRATIO, they are just entered for user reference. The energies must be sorted in
ascending order. The energies and statistical weights are stored in global variables
E(i) and G(i), respectively.

\begin{tabbing}
\qquad \=\qquad \=\qquad  \kill
\# \\
9 \\
1 \> 2 \> 8.865E-5  \\
1 \> 3 \> 1.275E-10 \\
2 \> 3 \> 1.772E-5  \\
1 \> 4 \> 6.535E-3  \\
2 \> 4 \> 2.111E-3  \\
3 \> 4 \> 6.388E-7  \\
1 \> 5 \> 2.945E-4  \\
2 \> 5 \> 7.909E-2  \\
4 \> 5 \> 1.124E0
\end{tabbing}
 
This block contains the transition probabilities. The number at the beginning is the number
of transitions listed. The first column is the lower level index 
$j$ for the transition, next are the higher level index $i$ and the transition probability
$A_{ij}$ (in s$^{-1}$). The transition probabilities are stored in the global variable A(i,j).
As the transition probabilities are read, INITPOPRATIO automatically sets the Einstein
coefficients by eq. (\ref{eq:EinsteinB}) and stores them in the global variable B(i,j).
There is no need to enter null transition probabilites, since transition probabilities not
entered are assumed to be zero.

\begin{tabbing}
\bigq \=\qquad \=\qquad \=\qquad  \kill
\# \\
135 \\
76794.798 \> 3 \> 1 \> 3.30E+08 \\	
76794.798 \> 3 \> 2 \> 1.97E+08	\\
76794.798 \> 3 \> 3 \> 6.54E+07	\\
96225.049 \> 3 \> 1 \> 9.31E+07	\\
96225.049 \> 3 \> 2 \> 5.56E+07	\\
96225.049 \> 3 \> 3 \> 1.85E+07	\\
97488.378 \> 3 \> 1 \> 1.99E+06	\\
97488.378 \> 3 \> 2 \> 2.97E+07	\\
97488.378 \> 3 \> 3 \> 3.95E+07	\\
97488.448 \> 5 \> 1 \> 1.79E+07	\\
97488.448 \> 5 \> 2 \> 5.35E+07 \\
...................................................
\end{tabbing}

These are the transitions involving the $\mu$ higher-lying levels of section 
\ref{section:fluorescence} and the NGROUND ground term levels. The number at the beginning is
the number of transitions listed. The first column is the energy relative to the ground level
(in cm$^{-1}$) of the higher-lying level, the second is its statistical weight, the third is
the lower-lying level index ($\leq$NGROUND) for the transition and the fourth is the transition
probability (in s$^{-1}$). The energy levels must be entered in ascending order.
If some higher-lying level was already loaded in by INITPOPRATIO in the fourth block (and
therefore will be explicitly included in the system of statistical equilibrium equations),
then it will not be taken into account when POPULATIONRATIO calculates the indirect excitation
rates, GAM(i,j). However its contribution will be added in TOTGAM(i,j), in case the user wishes
to assess it.

\# \break
4

This is the number of collision partners.
This value is stored in the global variable NPARTNERS and is the dimension of the collisional
data type vector CD(i).

Next we have one block for each one of the collisional partners:

\begin{tabbing}
\qquad \=\qquad \=\bigq \=\bigq \=\bigq \=\bigq \=\bigq \=\bigq \=\bigq  \kill
\# \\
electron \\
5 \\
10 \\
3 \> 7 \\
\> \> 50. \> 100. \> 200. \> 500. \> 1000. \> 2000. \> 3000. \\
2 \> 1 \> 8.34E-04 \> 1.26E-03 \> 1.79E-03 \> 2.58E-03 \> 3.35E-03 \> 4.61E-03 \> 5.92E-03 \\	
3 \> 1 \> 3.23E-04 \> 5.00E-04 \> 7.34E-04 \> 1.12E-03 \> 1.49E-03 \> 2.08E-03 \> 2.64E-03 \\	
3 \> 2 \> 2.74E-07 \> 8.92E-07 \> 2.78E-06 \> 1.05E-05 \> 2.38E-05 \> 4.62E-05 \> 6.98E-05 \\	
.......................................................................................................................................
\end{tabbing}

The first line is the collisional partner name, with a maximum of 10 characters long. This string
is stored in the global variable CD(i)\%PARTNER\_NAME. The second line is the index of the
highest-lying level that may be populated by collisions with this partner; it is stored in
CD(i)\%MAXLEVEL. The third line is the total number of transitions listed below; it is
stored in CD(i)\%NINTERPOL. For each transition we have a table of points containing the
collision rates (or Maxwellian-averaged collision stregths, if the partner name is 'ELECTRON')
as a function of the kinetic temperature. PopRatio interpolates this table of points by a
cubic-spline in log-log space (linear-log, in case the partner name is 'ELECTRON'). PopRatio
also uses eq. (\ref{eq:detailedbalance}) to calculate the inverse process rate.

In the next
line we have the partial number of transitions that follows and the number of data points 
entered in the tables. The next line lists the temperature values (in K) for the transitions
that follows. Then we have the collision rates (in cm$^3$s$^{-1}$, or Maxwellian-averaged
collision strengths), the first two values entered giving the $i$ and $j$ indexes for the
transitions ($i\rightarrow j$). If the partner name is electron then the order of these indexes
are irrelevant, otherwise $i$ must come first. This is repeated until the sum of the partial
number of transitions equals the total number of transitions listed. This is to allow greater
flexibility, since rates for the various transitions might be taken from different sources in
the literature and therefore have been calculated in different temperature values.

\begin{tabbing}
\qquad \=\qquad \=\bigq \=\quad \kill
\# \\
proton \\
3 \\
3 \\
3 \> 2 \\
\> \> 1.E30 \> 1.E30 \\
2 \> 1 \> 1.E-30 \> 1.E-30 \\
3 \> 1 \> 1.E-30 \> 1.E-30 \\
3 \> 2 \> 1.E-30 \> 1.E-30
\end{tabbing}

This is an example of how the user can implement an analytical fitting. The transitions
induced by collisions with protons are entered, but with an unphysical temperature
range. Since the temperature passed to POPULATIONRATIO will always lie outside this range, it
will call RATE\_EXTRAPOL that is then used to call OI\_PROTON, containing the analytical
fittings.

The last two blocks in OI.dat give the collision rates for transitions induced by collisions
with neutral hydrogen and neutral helium.

\subsection{Installing and running PopRatio}

\subsubsection{Setting the accuracy of floating point operations}

The user must set the accuracy of the floating point calculations by editing the WP kind
parameter in module PRECISION (see section \ref{section:Precision}).

There are only two options available, corresponding to the single/double precision Fortran
77's data types. This restriction applies because PopRatio uses F77 routines from the
BLAS, LAPACK and PPPACK libraries. The Fortran 90 interfaces to these routines will
automatically call the correct routines suitable to the selected precision.

The default precision is set to correspond to the {\bf double precision} Fortran 77 data type.

\subsubsection{Setting the radiation field}

The user must supply the radiation field he/she is interested in by editing the body of
function URAD in module POPRATIO.

For example, in order to include the FGAL and FUVB radiation fields defined in module
FIELDINTENSITIES the user should change the body of this function to:

\begin{tabbing}
\quad \=\quad \kill
urad = Fgal(lambda) \\
{\bf if} (z/=-1.0\_WP) {\bf then} \\
\> urad = urad + Fuvb(lambda,z,3) \\
{\bf end if}
\end{tabbing}

If the user does not intend to use any of the radiation fields defined in module
FIELDINTENSITIES then it does not need to be included in the project,
and the corresponding {\bf use} FieldIntensities statement may be deleted from module POPRATIO.

The default is a null radiation field:

urad = 0.0\_WP ;

\subsubsection{Writing the main program}

A minimal main program should include an {\bf open} statement specifying the atomic data
input file, a {\bf call} InitPopRatio statement loading in a model for the atom/ion being
considered and a {\bf call} FinishPopRatio statement to deallocate the dynamical variables,
freeing up RAM memory.

Typically between the calls to INITPOPRATIO and FINISHPOPRATIO we will have several calls to
POPULATIONRATIO under different physical conditions. After each call to POPULATIONRATIO
the rates for all the processes involved and the population ratios will be acessible via the
global variables given in table \ref{table:globalvariables}.

The calculation will proceed faster in case the physical conditions are not very different
from the physical conditions in last call to POPULATIONRATIO. If the same value for the
kinetic temperature T is passed in two successive calls to POPULATIONRATIO, then the
collision rates CD(i)\%q(i,j) will not need to be re-computed. Similarly, if we have
the same values for Z, BETA and F parameters, the radiation field energy densities U(i,j)
and the indirect excitation rates GAM(i,j) will remain unchanged.

Finally, a note of caution: if the user is interested in the collision rate $q_{ij}$ by a given
collisional partner, then he/she should not call POPULATIONRATIO with its density set to zero,
since the collision rates will not be calculated in this case.

In the next section we give some examples of main programs to tackle common applications.

\section{Applications and examples of using PopRatio}

\label{section:testcases}

\subsection{Testcase \#1}

In this example we calculate the population ratios of the $^3$P$_J$ ground levels of
atomic carbon as a function of neutral hydrogen density. These population ratios are
a useful density estimator \cite{Keenan} and are needed
to interpret ratios of UV C I fine structure absorption lines observed in damped
Lyman-$\alpha$ QSO absorbers \cite{Songaila,Ge,Roth}, for example.

We take into account collisions with several collisional partners - electron, proton,
neutral hydrogen, para/ortho molecular hydrogen, neutral helium -, tying their densities to
the neutral hydrogen density $n_{H^0}$ according to:

\begin{eqnarray}
n_e           & = & 10^{-4} n_{H^0} \nonumber \\
n_p           & = & 0 \nonumber \\
n_{p-H_2}  & = & \frac{1}{4} n_{H^0} \\
n_{o-H_2} & = & \frac{3}{4} n_{H^0} \nonumber \\
n_{He^0}        & = & \frac{1}{10} n_{H^0} \nonumber \  .
\end{eqnarray}

We also take into account the effect of CMBR at a redshift $z=1$ and the radiation field of
the Galaxy.

In the high density limit, function LTEDEV is called to show that the population ratios differ
from LTE conditions by less than 4\% .

To run this testcase the user should include the radiation field of the Galaxy in the body
of function URAD in module POPRATIO:

urad = Fgal(lambda) 

\subsection{Testcase \#2}

Now we want to determine the neutral hydrogen density from the population ratio (deduced from
the observed column density ratios of UV lines in the spectrum) assuming the same physical
conditions as in the previous example.

This example exploits the fact that the diagnosis curve traced in the first example is
a monotonically increasing function of density.

To run this testcase the user should include the radiation field of the Galaxy in the body
of function URAD in module POPRATIO:

urad = Fgal(lambda) 

\subsection{Testcase \#3}

This example reads in the neutral hydrogen density determined in the previous example and,
assuming the same physical conditions as above, determines the rates for all the processes
involved in order to access what are the most important excitation mechanisms.

The results show that the most important excitation mechanism are collisions by neutral
hydrogen, with the CMBR making a significant contribution for the $^3$P$_1$ level excitation.

To run this testcase the user should include the radiation field of the Galaxy in the body
of function URAD in module POPRATIO:

urad = Fgal(lambda) 

\subsection{Testcase \#4}

We now illustrate how to use PopRatio to calculate collisionally excited emission line
intensity ratios.

As an example we take the UV intercombination multiplet 
2s2p$^2\ ^4$P $\rightarrow$ 2s$^2$2p$\ ^2$P$^o$ of C II at 2325 \AA .
Intensities ratios of lines belonging to this multiplet may serve as a useful indicator of
electronic densities in the range $10^7 \leq n_e \leq 10^{10} \ \mathrm{cm}^{-3}$
\cite{Nussbaumer,Lennon}.
These lines have been observed in the spectra of a variety of astronomical objects:
planetary nebulae \cite{Stencel}, giant stars \cite{Judge,Carpenter},
symbiotic stars \cite{Nussbaumer86} and in the solar chromosphere and transition region
\cite{Doschek}.

The emissivity of a line is given by \cite{Osterbrock}:

\begin{equation}
\epsilon \left( \lambda_{ij} \right) = n_i \ A_{ij} \ h \nu_{ij} \ .
\end{equation}

We calculate the following emissivity ratios as a function of electronic density:

\begin{eqnarray}
R_1 & = & \frac {\epsilon \left( ^4\mathrm{P}_{5/2}\rightarrow{^2\mathrm{P}^o_{3/2}} \right)} 
		{\epsilon \left( ^4\mathrm{P}_{1/2}\rightarrow{^2\mathrm{P}^o_{3/2}} \right)}
					\nonumber \\
R_2 & = & \frac {\epsilon \left( ^4\mathrm{P}_{5/2}\rightarrow{^2\mathrm{P}^o_{3/2}} \right)} 
		{\epsilon \left( ^4\mathrm{P}_{3/2}\rightarrow{^2\mathrm{P}^o_{3/2}} \right)} \\
R_3 & = & \frac {\epsilon \left( ^4\mathrm{P}_{1/2}\rightarrow{^2\mathrm{P}^o_{1/2}} \right)} 
		{\epsilon \left( ^4\mathrm{P}_{3/2}\rightarrow{^2\mathrm{P}^o_{3/2}} \right)}
					\nonumber \ .
\end{eqnarray}

We take into account collisions by electrons and protons and fluorescence induced by a black
body radiation field of temperature $T_*=4000$K attenuated by a geometric dilution factor
$f=0.5$ (which might be representative of a stellar chromosphere).

In the high density limit function LTEDEV is called to show that the level populations differ
from LTE conditions by less than 0.4\%.

To run this testcase the user must include the black body radiation field in the body
of function URAD in module POPRATIO:

urad = Planck(4.0E3\_WP,c/lambda*1.0E8\_WP)

\subsection{Testcase \#5}

In this example we illustrate how PopRatio may be used to calculate the cooling rate due
to collisional excitation of low-lying levels of a given atom or ion.

As an example we take the ion Fe$^+$, because iron is an astrophysically abundant element.
The ion Fe$^+$ may be the dominating ionization stage in low ionization regions, and
may be an important gas coolant \cite{Nuss88}.

The cooling rate is given by \cite{Osterbrock}:

\begin{equation}
L \left( T \right) = \sum_{i=2}^{i_{\mathrm{max}}} \sum_{j<i} \ n_i \ A_{ij} \  h \nu_{ij} \ . 
\end{equation}

Rewriting this in terms of the total Fe$^+$ density $n_{\mathrm{Fe}^+} = \sum_i n_i$
and the population ratios $X_i=\frac{n_{i+1}}{n_1}$:

\begin{equation}
\frac{L(T)}{n_{\mathrm{Fe}^+}} = {( 1 + \sum_i X_i )}^{-1}
\sum_{i=2}^{i_{\mathrm{max}}} \sum_{j<i} X_{i-1} A_{ij} \left( E_i - E_j \right) hc \ ,
\label{eq:cooling} 
\end{equation}

with the energies $E_i$ expressed in cm$^{-1}$. The Fe$^+$ density will depend on its
fractional abundance and on the iron elemental abundance:

\begin{equation}
n_{\mathrm{Fe}^+} = \frac{n_{\mathrm{Fe}^+}}{n_{\mathrm{Fe}}} \ 
\frac{n_{\mathrm{Fe}}}{n_H} \ n_H
\end{equation}

Here we calculate the {\it cooling function}, defined as the right hand side of
eq. (\ref{eq:cooling}).

Since Fe$^+$ has a complicated electronic structure,
several levels must be taken into account in the calculation. We employ a 16-level model ion,
allowing us to calculate the cooling function for electronic densities as high as
10$^4$ cm$^{-3}$. 

To run this testcase the user does not need to modify function URAD, since no fluorescence
transitions are loaded in.

\section{Concluding remarks}

We have described a program to calculate atomic level population ratios. The code may find
a wide range of applicability in astronomical problems, as we have illustrated in the
testcases.

\begin{ack}

We would like to thank F. Haardt for providing us an electronic version of his UVB field.
A.I.S. acknowledges finantial support by the Brazilian agency FAPESP, under contract 
No. 99/05203-8. This work is partially supported by CNPq (304077/77-1) and
PRONEX/FINEP (41.96.0908.00).

\end{ack}

\end{document}